\documentstyle[aps,amstex,epsfig,fleqn]{revtex}
\newcommand{\erf}{\operatorname{erf}}
\def\hbar{{\mathchar'26\mskip-9muh}}

\begin{document}

\title{Quantum Monte Carlo Method for Attractive Coulomb Potentials}

\author{J.S. Kole and H. De Raedt}

\address{Institute for Theoretical Physics and
Materials Science Centre,\\
University of Groningen, Nijenborgh 4,
NL-9747 AG Groningen, The Netherlands}

\date{DRAFT: \today}
\maketitle
\begin{abstract}
Starting from an exact lower bound on the imaginary-time propagator,
we present a Path-Integral Quantum Monte Carlo method that
can handle singular attractive potentials.
We illustrate the basic ideas of this Quantum Monte Carlo algorithm
by simulating the ground state of hydrogen and helium.

\end{abstract}

\section{Introduction}
Quantum Monte Carlo (QMC) simulation is a powerful method for computing
the ground state and non-zero temperature properties of quantum many-body
systems\cite{SCHCEP,HDRVDL}.
There are two fundamental problems that limit the application
of these methods. The first and most important is the minus-sign problem
on which we have nothing to say in this paper, see however\cite{HDRFRI,HDRFET}.
The second problem arises if one would like to simulate
systems with attractive singular potentials, the Coulomb interaction being the prime example.
The purpose of this paper is to present an approach that solves the latter problem,
in a form that fits rather naturally in the standard Path Integral QMC (PIQMC)
approach and leaves a lot of room for further systematic improvements.

Let us first recapitulate the basic steps of the procedure to set up a PIQMC
simulation. Writing $K$ and $V$ for the kinetic and potential energy respectively
the first step is to approximate the imaginary-time propagator
by a product of short-time imaginary-time propagators.
The standard approach is to invoke the Trotter-Suzuki formula\cite{Lie,SUZUKIone}

\begin{equation}
e^{-\beta(K+V)}
=\lim_{m\rightarrow\infty}
\left( e^{-\beta K/m}e^{-\beta V/m)}\right)^m,
\end{equation}
to construct a sequence of systematic approximations $Z_m$
to the partition function $Z$\cite{SUZUKIone,HDRADL}:

\begin{eqnarray}
Z&=&\hbox{\bf Tr} \exp(-\beta H) =\lim_{m\rightarrow\infty} Z_m
\\ %\nonumber
Z_m&=&\int dr_1\cdots dr_m
\prod_{n=1}^m
\langle r_n|e^{-\beta K/m}|r_{n+1}\rangle e^{-\beta V(r_{n+1})/m},
\end{eqnarray}
where $r_{m+1}=r_1$ and use has been made
of the fact that the potential energy
is diagonal in the coordinate representation.
Taking the limit $m\rightarrow\infty$, (3) yields the Feynman
path integral\cite{Feyhib} for a system with Hamiltonian $H=K+V$.
Expression (3) is the starting point for PIQMC simulation.

In the case the attractive Coulomb interaction, it is easy
to see why the standard PIQMC approach fails.
Let us take the hydrogen atom as an example.
The Hamiltonian reads

\begin{equation}
H=-\frac{\hbar^2}{2M}\nabla^2-\frac{q^2}{r},
\end{equation}
where $q$ denotes the charge of the electron and $M=m_e/(1+m_e/m_p)$,
$m_e$ ($m_p$) being the mass of the electron (proton).
Replacing the imaginary-time free-particle propagator in (3) by its
explicit, exact expression

\begin{equation}
\langle r|e^{-\beta K/m}|r^\prime\rangle=
\left(\frac{mM}{2\pi \beta\hbar^2 }\right)^{3/2}
\exp\left(-\frac{mM|x-x'|^2}{2\beta\hbar^2}\right),
\end{equation}
we obtain

\begin{equation}
Z_m=\left( \frac{mM}{2\pi \beta\hbar^2}\right)^{3m/2} \int dr_1\cdots dr_m
\exp\left[ - \frac{mM}{2\beta\hbar^2} \sum_{n=1}^{m} (r_n-r_{n+1})^2\right]
\exp\left[ + \frac{\beta q^2}{m} \sum_{n=1}^{m} \frac{1}{r_n} \right].
\end{equation}
PIQMC calculates ratio of integrals such as (6) by using a Monte Carlo
procedure to generate the coordinates $\{r_1,\ldots,r_m\}$.
The integrand in (6) serves as the weight for the importance sampling process.
As the latter tends to maximize the integrand, it is clear that because
of the factors $\exp\left( + \beta q^2 m^{-1}r_n^{-1} \right)$,
the points $\{r_1,\ldots,r_m\}$ will, after a few steps, end up very close
to the origin. In the case of a singular, attractive potential
importance sampling based on (6) fails. Using instead of the simplest
Trotter-Suzuki formula (1) a more sophisticated one \cite{Suzukithree}
only makes things worse because these hybrid product formulae contain
derivatives of the potential with respect to the coordinates.

The problem encountered in setting up a PIQMC scheme for models with
a singular, attractive potential is just a signature of the fundamental
difficulties that arise when one tries to define the Feynman
path integral for the hydrogen atom\cite{Kleinertbook}.
The formal solution to this problem is known\cite{Kleinertbook,Duru}.
It is rather complicated and not easy to incorporate in a PIQMC simulation.

In spirit the method proposed in this paper is similar to the one used
to solve the hydrogen path integral: Use the quantum fluctuations
to smear out the singularity of the potential.
Mathematically we implement this idea by applying
Jensen's inequality to the propagator\cite{Symanzik}.
Applications of the Feynman path-integral formalism are often based on
a combination of Jensen's inequality and a variational approach\cite{Feyhib,Kleinertbook}
so it is not a surprise that similar tricks may work for PIQMC as well.

The paper is organized as follows. In Section 2 we give a simple
derivation of an exact lower bound on the imaginary-time propagator.
This inequality naturally defines a sequence of systematic approximations $\hat Z_m$
to the partition function.
Although each $\hat Z_m$ looks very similar to $Z_m$, the former
can be used for PIQMC with attractive, singular potentials.
For pedagogical reasons, in Section 3 we illustrate the approach
by presenting an analytical treatment of the harmonic oscillator.
In Section 4 we give the explicit form of the approximate
propagator for the attractive Coulomb potential and present PIQMC
results for the ground state of the hydrogen and helium atom.

\section{Lower bound on the propagator}
Consider a system with Hamiltonian $H=K+V$ and a complete set
of states $\{ |x\rangle \}$ that diagonalizes the hermitian
operator $V$.
In the case that $V$ contains a singular attractive part we replace
$V=\lim_{\epsilon\rightarrow0} V_\epsilon$ by a regular $V_\epsilon(x)>-\infty$
and take the limit $\epsilon\rightarrow0$ at the end of the calculation.
Using the Trotter-Suzuki formula we can write

\begin{eqnarray}
\langle x|e^{-\tau(K+V_\epsilon)}|x'\rangle
&=&
\lim_{m\rightarrow\infty}\langle x|\left( e^{-\tau K/m}e^{-\tau V_\epsilon/m}\right)^m|x'\rangle,
\\
& = &
\lim_{m\rightarrow\infty}
\int dx_1\cdots dx_n\prod_{i=1}^{m}\langle x_i|e^{-\tau K/m}|x_{i+1}\rangle e^{-\tau V_\epsilon(x_i)/m},
\\
& = &
\lim_{m\rightarrow\infty}
\frac{
\int dx_1\cdots dx_n\prod_{i=1}^{m}\langle x_i|e^{-\tau K/m}|x_{i+1}\rangle e^{-\tau V_\epsilon(x_i)/m}
}{
\int dx_1\cdots dx_n\prod_{i=1}^{m}\langle x_i|e^{-\tau K/m}|x_{i+1}\rangle
}
\int dx_1\cdots dx_n\prod_{i=1}^{m}\langle x_i|e^{-\tau K/m}|x_{i+1}\rangle
.
\end{eqnarray}
If $\langle x|e^{-\tau K}|x^\prime\rangle \ge 0$ for all $\tau$, $x$ and $x^\prime$,
the function

\begin{equation}
\rho(\{x_i\}) = \left.
\prod_{i=1}^{m}\langle x_i|e^{-\tau K/m}|x_{i+1}\rangle\right/
\int dx_1\cdots dx_n\prod_{i=1}^{m}\langle x_i|e^{-\tau K/m}|x_{i+1}\rangle,
\end{equation}
is a proper probability density. Clearly (10) is of the form
$\int dx_1\cdots dx_n \rho(\{x_i\})f(\{x_i\})$ so that
we can apply Jensen's inequality

\begin{equation}
\int dx_1\cdots dx_n  \rho(\{x_i\})e^{g(\{x_i\})}
\ge
\exp\left(\int dx_1\cdots dx_n  \rho(\{x_i\}) g(\{x_i\})\right),
\end{equation}
and obtain

\begin{eqnarray}
\langle x|e^{-\tau(K+V_\epsilon)}|x'\rangle
&\ge&
\langle x|e^{-\tau K}|x'\rangle
\lim_{m\rightarrow\infty}
\exp \left(-\frac{\tau}{m}\sum_{i=1}^{m}
 \int dx_1\cdots dx_m
\frac{V_\epsilon(x_i)\prod_{n=1}^{m}\langle x_n|e^{\tau K/m}|x_{n+1}\rangle}{\langle x|e^{-\tau K}|x'\rangle}
  \right),
\\
&\ge&
\langle x|e^{-\tau K}|x'\rangle
\lim_{m\rightarrow\infty}
\exp \left( -\frac{\tau}{m}\sum_{i=1}^{m}
 \int dx_i
\frac{\langle x|e^{\tau K/m}|x_i\rangle V_\epsilon(x_i)
\langle x_i|e^{\tau K/m}|x^\prime\rangle }{\langle x|e^{-\tau K}|x'\rangle}
  \right).
\end{eqnarray}
For $m\rightarrow\infty$, the sum over $n$ can be replaced by an integral over imaginary time.
Finally we let $\epsilon\rightarrow0$ and obtain\cite{Symanzik}

\begin{eqnarray}
\langle x|e^{-\tau(K+V)}|x'\rangle
&\ge&
\langle x|e^{-\tau K}|x'\rangle
\exp\left\{
-\int^{\tau}_{0}du \frac{
\langle x|e^{-uK}V e^{-(\tau-u)K}|x'\rangle}{\langle x|e^{-\tau K}|x'\rangle}\right\}.
\end{eqnarray}
Note that l.h.s of (14) reduces to the standard, symmetrized Trotter-Suzuki formula
approximation\cite{hansbart,suzukitwo}
if we replace the integral over $u$ by a two-point trapezium-rule approximation.
This replacement also changes the direction of inequality
as can been seen directly from the upperbound\cite{Symanzik}

\begin{eqnarray}
\langle x|e^{-\tau(K+V)}|x'\rangle
&\le&
\langle x|e^{-\tau K}|x'\rangle
\exp\left\{
-\int^{\tau}_{0}du
\ln\left(\frac{ \langle x|e^{-uK}e^{-\tau V} e^{-(\tau-u)K}|x'\rangle }{\langle x|e^{-\tau K}|x'\rangle}\right)
\right\}
\le
\langle x|e^{-\tau K}|x'\rangle e^{-\tau V(x)}
.
\end{eqnarray}

Expression (14) can be used to define a new type of approximant to the partition function
namely

\begin{eqnarray}
\hat Z_m=\left( \frac{M}{2\pi \tau\hbar^2}\right)^{3m/2} \int dr_1\ldots dr_m
\prod_{n=1}^{m} \exp\left[ - \frac{M}{2\tau\hbar^2} (r_n-r_{n+1})^2
-\int^{\tau}_{0}du \frac{
\langle r_n|e^{-uK}V e^{-(\tau-u)K}|r_{n+1}\rangle}{\langle r_n|e^{-\tau K}|r_{n+1}\rangle}
\right]
.
\end{eqnarray}
where $\tau=\beta/m$.
The simplest approximant $\hat Z_1$ corresponds to the Feynman's variational approximation
to the full Feynman path integral\cite{Feyhib,Kleinertbook}.
The main difference between (3) and (16) is that the bare potential $e^{-\tau V(x)}$
is replaced by an effective potential that is obtained by convoluting
the bare potential and free-particle propagators $e^{-uK}$ and $e^{-(\tau-u)K}$.
Convolution smears out singularities. As we show below, in the case of the attractive
Coulomb interaction expression (14) is finite, for any choice of $x$ and $x^\prime$.
For the approximants $\hat Z_m$ to be useful in PIQMC, it is necessary that
the integral over $u$ can be done efficiently. In the next two sections we
show how this can be done.

\section{Illustrative example}
It is instructive to have at least one example for which
the details can be worked out analytically, without actually using PIQMC.
Not surprisingly this program can be carried out for the harmonic
oscillator. For notational convenience we will consider the one-dimensional
model Hamiltonian $H=K+V$, with $K=-(\hbar^2/2M) d^2/dx^2$
and $V=M\omega^2 x^2$. Calculating the matrix element
$\langle x|e^{-uK}V e^{-(\tau-u)K}|x'\rangle$ in (16)
is a straightforward excercise in perfoming Gaussian integrals\cite{note1}.
We obtain

\begin{eqnarray}
\hat Z_m =
\left(\frac{mM}{2 \pi\beta\hbar^2}\right)^{m/2}
     \int dx_1 \ldots dx_{m} \prod^{m}_{n=1}\exp\left[-
          \frac{mM}{2\beta\hbar^2}(x_n-x_{n+1})^2-
          \frac{\beta M\omega^2}{6m}(x_{n}^{2}+x_{n+1}^{2}+x_{n}x_{n+1}+
          \frac{\beta\hbar^2}{2mM} )
          \right]
.
\end{eqnarray}
The integrand in (17) is a quadratic form and can be diagonalized
by a Fourier transformation with respect to the index $n$.
Evaluation of the resulting Gaussian integrals yields

\begin{eqnarray}
\hat Z_m =
2^{-m/2}
          \exp\left(-\frac{\beta^2\hbar^2\omega^2}{12m}\right)
          \prod_{n=0}^{m-1}\left[1+
          \frac{\beta^2\hbar^2\omega^2}{3m}-\left(1-
          \frac{\beta^2\hbar^2\omega^2}{6m}\right)
          \cos\left(\frac{2\pi n}{m}\right)\right]^{-1/2}.
\label{eqn2}
\end{eqnarray}
Taking the partial derivative of $-\ln\hat Z_m$ with respect to $\beta$
gives the corresponding approximation to the energy:
\begin{equation}
\hat E_m = \frac{\beta\hbar^2\omega^2}{6m}\left[1+ \sum_{n=0}^{m-1}
      \frac{2+\cos\left(2\pi n/m\right)}{1-\cos\left(2\pi n/m\right)
          +\beta^2\hbar^2\omega^2\left(2+\cos(2\pi n/m)\right)/6m
          }\right]
          .
\end{equation}
For comparison, if we use of the standard Trotter-Suzuki formula we obtain\cite{HDRADL}
\begin{equation}
E_m = \frac{\beta\hbar^2\omega^2}{2m^2}\sum_{n=0}^{m-1}\frac{1}
     {1-\cos\left(2\pi n/m\right)+\beta^2\hbar^2\omega^2/2m^2}
\end{equation}

In Table 1 we present numerical results obtained from (19) and (20)
and compare with the exact value of the energy
$E=(\hbar\omega/2)\text{coth}(\beta\hbar\omega/2)$).
Note that the average of the two approximations, i.e. $(\hat E_m+E_m)/2$, is remarkably
close to the exact value $E$, an observation for which we have no mathematical
justification at this time.

\begin{table}[h]
\caption{Numerical results for the exact energy of the harmonic oscillator ($E$),
and approximations based on (19) ($\hat E_m$) and (20) ($E_m$). We use units
such that $\hbar\omega=1$ and $\beta$ is dimensionless.}
\begin{tabular}[h]{rrccc}
$\beta$ & m & $E_m$ & E & $\hat E_m$ \\
\hline
1 &    1 & 1.00000 & 1.08198 & 1.16668 \\
  &   10 & 1.08101 & 1.08198 & 1.08292 \\
  &   50 & 1.08194 & 1.08198 & 1.08202 \\
  &  100 & 1.08197 & 1.08198 & 1.08199 \\
  &  500 & 1.08198 & 1.08198 & 1.08198 \\
\hline
5 &    1 & 0.20000 & 0.50678 & 1.03333 \\
  &   10 & 0.49199 & 0.50678 & 0.51938 \\
  &   50 & 0.50617 & 0.50678 & 0.50694 \\
  &  100 & 0.50678 & 0.50678 & 0.50679 \\
  &  500 & 0.50678 & 0.50678 & 0.50679 \\
\hline
10&    1 & 0.10000 & 0.50005 & 1.76667 \\
  &   10 & 0.44273 & 0.50005 & 0.54316 \\
  &   50 & 0.49757 & 0.50005 & 0.50234 \\
  &  100 & 0.49942 & 0.50005 & 0.50064 \\
  &  500 & 0.50002 & 0.50005 & 0.50007 \\
\end{tabular}
\label{table1}
\end{table}

\section{Attractive Coulomb Potential}

As a second example we will consider a neutral system consisting of two electrons with opposite spin and a
nucleus. The Hamiltonian reads\cite{schiff,baym}

\begin{equation}
H=-\frac{\hbar^2}{2M_1}\nabla_1^2 - \frac{\hbar^2}{2M_2}\nabla_2^2
-\frac{q^2}{|r_1|}-\frac{q^2}{|r_2|}+\frac{2q^2}{|r_1-r_2|},
\end{equation}
where the vectors $r_1$ and $r_2$ describe the position of the two electrons,
with the nucleus placed in the origin.
It is convenient to introduce the notation
$K_i=-D_i^{\phantom{2}}\nabla_i^2$, $D_i=\hbar^2/2M_i$,
$V_i=V(r_i)$, $V_{12}=V(r_1-r_2)$,
and $V(r)=q^2/|r|$, for $i=1,2$.
Application of inequality (14) requires the evaluation of

\begin{eqnarray}
I(r_1,r_2,r'_1,r'_2) &=&- \frac{\int^{\tau}_{0}du \langle r_1 r_2|e^{-u(K_1+K_2)}(V_1+V_2-2V_{12})
     e^{-(\tau-u)(K_1+K_2)}|r^\prime_1r^\prime_2\rangle}{
     \langle r_1r_2|e^{-\beta (K_1+K_2)}|r^\prime_1r^\prime_2\rangle}
\\ \nonumber
&=&
-\frac{\int^{\tau}_{0}du\langle r_1|e^{-u K_1} V_1
     e^{-(\tau-u) K_1}|r_1'\rangle}
     {\langle r_1|e^{-\tau K_1)}|r_1'\rangle}
-\frac{\int^{\tau}_{0}du\langle r_2|e^{-u K_2} V_2
     e^{-(\tau-u) K_2}|r_2'\rangle}
     {\langle r_2|e^{-\tau K_2}|r_2'\rangle}
\\ \nonumber
&&+2\frac{\int^{\tau}_{0}du \langle r_1r_2|e^{-u(K_1+K_2)}V_{12}
     e^{-(\tau-u)(K_1+K_2)}|r'_1r'_2\rangle}{
     \langle r_1r_2|e^{-\tau (K_1+K_2)}|r^\prime_1r^\prime_2\rangle},
\end{eqnarray}
where we made use of the fact that $[K_1,V_2]=[K_2,V_1]=0$.
It is sufficient to consider the last term of (22).
Inserting a complete set of states for both particles we obtain

\begin{eqnarray}
I_{12}(r_1,r_2,r'_1,r'_2)
&=&\frac{\int^{\tau}_{0}du \int dr_1''\int dr_2''\langle r_1r_2|e^{-u(K_1+K_2)}|r_1''r_2''\rangle
V(r_1''-r_2'')
     \langle r_1''r_2''|e^{-(\tau-u)(K_1+K_2)}|r'_1r'_2\rangle}{
     \langle r_1r_2|e^{-\tau (K_1+K_2)}|r^\prime_1r^\prime_2\rangle}
     .
\end{eqnarray}
Inserting the explicit expression for the free-particle propagator (5), a straightforward
manipulation of the Gaussian integrals in (23) gives

\begin{eqnarray}
I_{12}(r_1,r_2,r'_1,r'_2,D)
= \int_0^{\tau} du\int dr% &&
\left(\frac{\tau}{4\pi u(\tau-u)D}\right)^{3/2}
V(r)
%\\ \nonumber && \times
\exp\left\{
-\frac{[\tau r-(\tau-u)(r_1-r_2)-u (r_1'-r_2')]^2}
      {4u\tau(\tau-u) D}\right\},
\end{eqnarray}
where $D=D_1+D_2$

In the case of the Coulomb potential, the integral over $r$ can be evaluated analytically
by changing to spherical coordinates. The remaining integral over $u$ is calculated
numerically. In practice, it is expedient to replace the integration over $u$ by
an integration over an angle. An expression that is adequate for numerical
purposes is

\begin{eqnarray}
I_{12}(r_1,r_2,r'_1,r'_2,D)=2\tau q^2\int_0^{\pi/2} d\phi
\frac
{
\erf\left[(4\tau D)^{-1/2}|(r_1-r_2)\tan\phi  + (r_1'-r_2')\cot\phi|\right]
}{
|(r_1-r_2)\tan\phi  + (r_1'-r_2')\cot\phi|
}.
\end{eqnarray}
It is easy to check that $I_{12}(r_1,r_2,r'_1,r'_2,D)$ is finite.
The expressions for the first and second
contributions in (22) can be obtained from (25) by putting ($D_2$, $r_2$, $r'_2$)
and ($D_1$, $r_1$, $r'_1$) equal to zero, i.e.
$I_1(r_1,r'_1,D_1)=I_{12}(r_1,0,r'_1,0,D_1)$ and
$I_2(r_2,r'_2,D_2)=I_{12}(0,r_2,0,r'_2,D_2)$.

For the helium atom $M=M_1=M_2$, and the $m$-th approximant to the partition function reads

\begin{eqnarray}
\hat Z_m^{He}=\left( \frac{M}{2\pi \tau\hbar^2}\right)^{3m} \int &&dr_1\ldots dr_m
dr'_1\ldots dr'_m
 \exp\left\{ - \frac{M}{2\tau\hbar^2} \sum_{n=1}^{m}\left[(r_n-r_{n+1})^2 +(r'_n-r'_{n+1})^2\right]
\right\}
\\ \nonumber
&&\times\exp\left\{\tau\sum_{n=1}^{m}\left[
 I_1(r_n,r_{n+1},D_1)+I_2(r'_n,r'_{n+1},D_1)
 -2 I_{12}(r_{n}^{\phantom{'}},r_{n+1}^{\phantom{'}},r'_{n},r'_{n+1},2D_1)\right]
\right\}
,
\end{eqnarray}
whereas in the case of the hydrogen atom we have

\begin{eqnarray}
\hat Z_m^{H}=\left( \frac{M}{2\pi \tau\hbar^2}\right)^{3m/2} \int dr_1\ldots dr_m
 \exp\left\{- \frac{M}{2\tau\hbar^2} \sum_{n=1}^{m}(r_n-r_{n+1})^2 +\tau \sum_{n=1}^{m}I_1(r_n,r_{n+1},D_1)
\right\}
,
\end{eqnarray}
with $\tau=\beta/m$.
As the integrands in (26) and (27) are always finite, expressions (26) and (27) can be used
perform PIQMC simulations.

In the path integral formalism
the ground state energy is obtained by letting $\beta\rightarrow\infty$ and
$\beta/m\rightarrow0$, i.e. $E=\lim_{\beta\rightarrow\infty}\lim_{\beta/m\rightarrow0}\hat E_m$.
Of course, in numerical work, taking one or both these limits is impossible.
In Tables 2 and 3 we present numerical results of PIQMC estimates of
the ground state energy $E$ of the hydrogen and helium atom.
These results have been obtained from five statistically independent simulations
of 100000 Monte Carlo steps per degree of freedom each.
The systematic errors due to the discretization of the path integral are hidden in
the statistical noise.
The PIQMC procedure we have used is standard\cite{SCHCEP,HDRADL} except for a trick
we have used to improve the efficiency of sampling the paths, details of which are given
in the appendix.
%For hydrogen the experimental value is
%$E=-0.5$ \cite{schiff,baym},
%for helium the exact value is
%$E=-2.904$ \cite{schiff}, both in units of $e^2/a_0$ ($a_0=\hbar^2/Me^2$).
Although a ground state calculation pushes the PIQMC method to point of becoming
rather inefficient, the numerical results are in satisfactory agreement with the known values.

\section{Discussion}

We have show that is possible to perform PIQMC
simulations for quantum systems with attractive Coulomb potentials.
Instead of the conventional Trotter-Suzuki formula approach one can
use (16) to construct a path integral that is free of singularities.
In practice, a numerical calculation of the latter requires only
minor modifications of a standard PIQMC code.

The efficiency of the PIQMC method describe above can be improved
with relatively modest efforts.
Instead of using the free-particle propagator $K$,
we are free to pick any other model Hamiltonian $H_0$ for which
the matrix elements of $e^{-\tau H_0}$ are positive
and integrals involving these matrix elements are known analytically.
An obvious choice would be to take for $H_0$ a set of harmonic oscillators.
The matrix elements of $e^{-\tau H_0}$ are Gaussians and hence the
conditions used to derive (14) are satisfied.
If necessary the approximant $\hat Z_m$ can be improved
further by optimization of the parameters of the oscillators.
For $m=1$ this approach is identical to the variational method proposed
by Feynman and Kleinert\cite{FeyKle,Kle1,Kle2,Janke}
and independently by Giachetti and Tognetti\cite{Gia1,Gia2}.
Extending the PIQMC method in this direction is left
for future research.

\begin{table}[h]
\label{table2}
\caption{Path-integral Quantum Monte Carlo results for
the ground state energy of the hydrogen Hamiltonian, in units of $q^2/a_0$
($a_0=\hbar^2/Mq^2$).
The exact value is $E=-0.5$.}
\begin{tabular}[h]{ccc}
$\beta$ & $m$ & $\hat E_m^H$ \\
\hline
20 & 400 & -0.496 ($\pm$ 0.004)\\
20 & 800 & -0.503 ($\pm$ 0.005)\\
40 & 800 & -0.498 ($\pm$ 0.006)\\
\end{tabular}
\end{table}

\begin{table}[h]
\label{table3}
\caption{Path-integral Quantum Monte Carlo results for
the ground state energy of the helium Hamiltonian, in units of $q^2/a_0$.
%($a_0=\hbar^2/Mq^2$).
The experimental value is $E=-2.904$.}
\begin{tabular}[h]{ccc}
%\multicolumn{3}{c|}{Helium} &
$\beta$ & $m$ & $\hat E_m^{He}$ \\
\hline
10  &  400 & -2.84 ($\pm$ 0.02)\\
10  &  800 & -2.88 ($\pm$ 0.02)\\
10  & 1200 & -2.92 ($\pm$ 0.03)\\
\end{tabular}
\end{table}

\section*{Appendix}

In PIQMC the simplest mehod for sampling paths is to change
one degree of freedom at each Monte Carlo step.
Usually this is rather inefficient and one adds Monte Carlo
moves that make global changes of the path, e.g. moves that
resembles the classical motion.
In this appendix we present a more sophisticated scheme which
we found performed very well at very low temperature.
The basic idea is to change variables such that the kinetic energy
term in the path integral becomes a diagonal quadratic
form, i.e.

\begin{eqnarray}
\sum_{k=1}^m \left( x_k-x_{k+1}\right)^2=\sum_{k=2}^m y_k^2,
\end{eqnarray}
where $x_{m+1}=x_1$.
After some straightforward algebra one finds that
the transformation from the $\{x_i\}$ to the $\{y_i\}$
is given by

\begin{eqnarray}
y_k^2=\frac{m-k+2}{m-k+1}
\left( x_k-\frac{(m-k+1)x_{k-1}+x_{m+1}}{m-k+2}\right)^2
.
\end{eqnarray}
The expression for $x_k$ in terms of the $\{u_i\}$  reads

\begin{eqnarray}
x_k=y_1+\sum_{j=2}^k \frac{m-k+1}{m-j+1}\left(\frac{m-j+1}{m-j+2}
\right)^{1/2}y_j,\quad 1<k\le m,
\end{eqnarray}
with $x_1=y_1$.
From (30) we conclude that the computational work for making
a global change of the path (i.e. simultaneously changing
all $y_i$) is linear in $m$, hence optimal.
It is also clear that the variable $y_1$ plays the role
of the ``classical'' position. The variables $y_2,\ldots,y_m$
describe the quantum fluctuations.


\begin{thebibliography}{99}
\bibitem{note1}
This is the case for all $V(x)$ that are polynomial in $x$.
\end{thebibliography}
\end{document}